\documentclass[journal]{IEEEtran}
\usepackage{amsmath}
\usepackage{mathptmx}

\usepackage{hyperref}
\usepackage{xcolor}
\hypersetup{
     colorlinks   = true,
     linkcolor = black,
     urlcolor  = blue,
     citecolor = blue,
     anchorcolor = red
}

\usepackage{graphicx}
\usepackage{url}
\usepackage{textcomp}
\usepackage[compress]{cite}

\DeclareMathOperator{\Imag}{\mathrm{Im}}

\begin{document}
\title{Experimental Investigation of Stub Resonators Built in Plasmonic Slot Waveguides}
\author{Solmaz~Naghizadeh,
	Adeel~Afridi,
	Ongun~Ar{\i}sev,
	Aziz~Kara\c{s}ahin
	and \c{S}\"ukr\"u~Ekin~Kocaba\c{s}~\IEEEmembership{Member,~IEEE}
	\thanks{This work was supported by the Scientific and Technological Research Council of Turkey (TUBITAK) under Grant No: 112E247.} %
	\thanks{S.\ Naghizadeh was with the Department
		of Physics, Ko\c{c} University, Sar{\i}yer, Istanbul, TR34450 Turkey. A.\ Afridi, O.\ Ar{\i}sev and \c{S}.\ E.\ Kocaba\c{s} are with the Department of Electrical and Electronics Engineering at Ko\c{c} University (email: ekocabas@ku.edu.tr). A.\ Kara\c{s}ahin is with the Department of Electrical and Computer Engineering at the University of Maryland, College Park MD 20742 USA.}
	\thanks{Copyright (c) 2016 IEEE. Personal use of this material is permitted.  However, permission to use this material for any other purposes must be obtained from the IEEE by sending a request to pubs-permissions@ieee.org.} %
	}

\maketitle

\begin{abstract}
In this work, we focus on stub resonators embedded in plasmonic slot waveguides. The resonators have potential applications in optical interconnects and sensors. We fabricate the samples by electron beam lithography and lift-off. We use a scattering matrix based model to quantify the optical power output from the samples. We measure the properties of the resonators by coupling light in and out of the slot waveguides by optical antennas, making use of a cross-polarization based setup utilizing a supercontinuum source and a high numerical aperture objective lens operating in the telecom-wavelength range.  Our model agrees well with the measured data. Further development on the stub resonators can be made by using the methods in this paper. 
\end{abstract}


\section{Introduction}

\IEEEPARstart{C}{urrent} microprocessors have hit a speed limit due to energy dissipation constraints. Most of the dissipation is on the electrical wires used to form the circuits. Optical interconnects offer a viable way to reduce energy consumption by eliminating the loss on electrical wires and by enabling high data rates over large distances with high density pathways \cite{Miller2017}. Such an interconnect system requires a number of optoelectronic components such as waveguides, light sources, modulators and detectors. When these components are tightly integrated with the transistors, it becomes conceivable to transfer a bit at the sub-femtojoule level, surpassing electrical interconnects \cite{Miller2017}. Nanophotonic devices that take advantage of the optical properties of metals can have deep subwavelength dimensions, and they can provide both optical and electrical connectivity at the same time \cite{Kinsey2015}---these properties are very relevant for intimate integration with transistors. 

There are a number of different plasmonic waveguide proposals, with different propagation lengths and mode sizes \cite{Kinsey2015}. Of these proposals, the plasmonic slot waveguide \cite{Veronis2007a} is a very promising candidate for integration. Free-space \cite{Wen2011,Andryieuski2014} and fiber based \cite{Gao2016a} couplers, interference based optical logic gates \cite{Fu2012}, photodetectors \cite{Ly-Gagnon2012}, 
mode converters \cite{Geisler2013,Dai2014}, directional couplers \cite{Kriesch2013,Rewitz2014}, modulators \cite{Melikyan2014,Lee2014a} and light sources \cite{No2013,Huang2014} for the slot waveguide geometry have been demonstrated. The use of low-Q resonators that do not require active tuning can further improve the properties of light emitters, modulators and photodetectors without incurring an energy penalty \cite{Miller2017}. 
Stub based resonators built in plasmonic slot waveguides can provide small mode volumes and be more compact compared to ring and Fabry-Perot counterparts \cite{Bozhevolnyi2006}. Stub resonators have been experimentally investigated in U-shaped plasmonic channel waveguides \cite{Zhu2014} and in a hybrid plasmonic slot--photonic crystal waveguide geometry \cite{Chai2016}. Straight, L- and T-shaped unconnected nanoslot resonator properties were experimentally probed in \cite{Raza2016}. Recently, properties of 3D stub resonators based on plasmonic slot waveguides have been modeled using scattering matrix theory and verified via \textsc{fdtd} simulations \cite{Naghizadeh2017}. In this letter, we experimentally investigate stub resonators built in plasmonic slot waveguides and compare measurements with the predictions of the model in \cite{Naghizadeh2017}.

\begin{figure}[!t]
\centering
\includegraphics{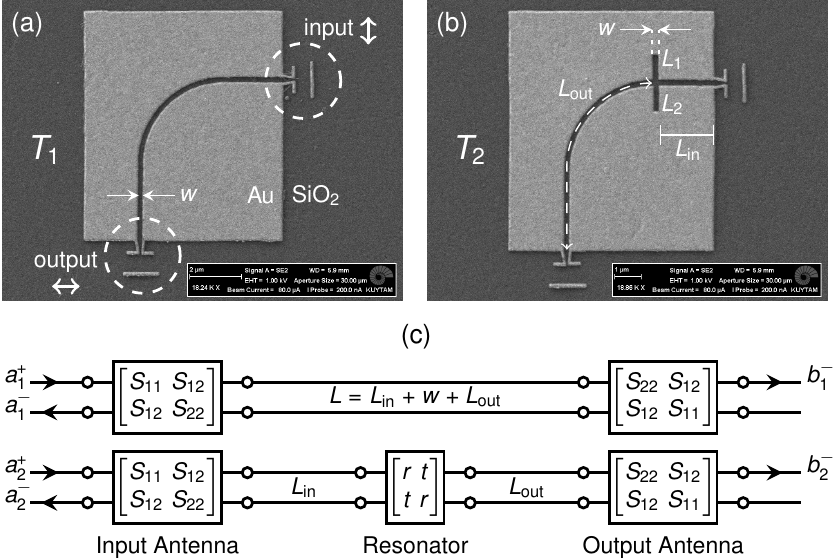}
\caption{SEM images of fabricated devices before SU8 coating for (a) samples without resonators, (b) samples with resonators. (c) Scattering matrix models for the devices.}\label{figsem}
\end{figure}

\section{Fabrication and Modeling}
Figure \ref{figsem} shows the \textsc{sem} images of the two types of samples used in the characterization of the stub resonators. We use gold (Au) as the metal layer and quartz (Si$\text{O}_2$) as the substrate. We employ dipole antennas \cite{Kriesch2013} to couple in and out of the slot waveguides. We build samples with no resonators in order to characterize the antenna--slot waveguide coupling rate and the modal properties of the slot waveguide [Fig.\ \ref{figsem}(a)]. We also build samples with double-stub resonators with stub lengths $L_1$ and $L_2$ embedded in the slot waveguide [Fig.\ \ref{figsem}(b)].  The slot width is fixed at $w=220$ nm and is also equal to the stub width. The distance between the resonator and the input (output) antenna is $L_\text{in}$ ($L_\text{out}$). Total length of the slot waveguide is the same in samples with and without resonators and is given by $L=L_\text{in}+w+L_\text{out}$.

The fabrication of the samples starts with spin coating of 4\% in anisole 495K \textsc{pmma} (Microchem) at 5000 rpm followed by 2\% in anisole 950K \textsc{pmma} at 5000 rpm to form a bi-layer resist on piranha cleaned quartz pieces. A thin (15 nm) aluminum layer is then evaporated to prevent the charging of the transparent substrate during e-beam lithography. A 100 kV e-beam lithography system (Vistec  EBPG5000plusES) is operated at the 850--950 \textmu{C}/$\text{cm}^2$ dose level. The aluminum layer is removed by a \textsc{tmah} based developer (Microchem AZ 726). Standard \textsc{pmma} development is made in 1:3 \textsc{mibk}:\textsc{ipa} solution for 1 min. 4 nm of Ti and 115 nm of Au are thermally evaporated at $2\times10^{-6}$ Torr pressure, at the rates 0.2 \AA/s and 0.5 \AA/s, respectively. These parameters have recently been found to be optimal for Au \cite{McPeak2015}. Lift-off procedure is done in heated ($\sim$50 \textdegree{C}) acetone, as well as room temperature acetone with ultrasonic agitation. Finally, a $\sim$500 nm layer of SU8 2000 resist (Microchem) is coated on the samples to improve the coupling rate of the dipole antenna to the slot waveguide \cite{Gao2016a} and to increase the propagation length of the slot waveguide mode.

We use a scattering matrix based model in order to quantify the transmitted optical power from the samples in Fig.\ \ref{figsem}(a)-(b). The input and output antennas have the same structure, therefore, they use the same set of matrix elements [Fig.\ \ref{figsem}(c) top]. However, compared to the input antenna, the waveguide is on the opposite side of the output antenna which leads to the $S_{11} \leftrightarrow S_{22}$ switch for the output antenna scattering matrix. The off-diagonal matrix elements are equal due to reciprocity. Samples with resonators have an additional scattering matrix composed of the reflection, $r$, and the transmission, $t$, coefficients of the double-stub resonator [Fig.\ \ref{figsem}(c) bottom]. 

The power transmission ratio of the devices without a resonator ($T_1)$ and those with a double-stub section ($T_2$) can be estimated based on the scattering matrix model via $T_i=\lvert b^-_i/a_i^+ \rvert^2$ for $i=1,2$ as shown in Fig.\ \ref{figsem}(c). In order to get closed form results for $T_i$, we convert the $S$ matrix for each section in the model into the corresponding $T$ matrix, multiply the $T$ matrices to obtain the overall $T$ matrix of the system (see e.g.\ \cite{Kocabas2008}) and from the inverse of the upper-left element of the overall $T$ matrix we arrive at
\begin{align}
T_1  &= |{S_{12}^2 e^{-i k L}}/({1-S_{22}^2 e^{-i 2 k L}})|^2, \label{eq.T1}\\
T_2  &= \left|\frac{t S_{12}^2 e^{-i k L^+}}{1-S_{22}^2 (t^2-r^2) e^{-i 2 k L^+}-2rS_{22} \cos(k L^-) e^{-i k L^+}}\right|^2,\label{eq.T2}
\end{align}
where $L^\pm=L_\text{in}\pm L_\text{out}$, and $k$ is the wave vector of the plasmonic slot waveguide mode.

A thorough analysis of the $r$ and $t$ coefficients of the double-stub geometry, as well as the characterization of the slot waveguide mode and its $k$ value are available in \cite{Naghizadeh2017}. We obtain $S_{12}$ and $S_{22}$ from \textsc{comsol} simulations. When obtaining $S_{12}$ we illuminate the antenna with a Gaussian beam and calculate the modal coupling coefficient. For $S_{22}$, the waveguide mode is injected towards the antenna and the reflection is measured.

\begin{figure}[!t]
	\centering
	\includegraphics{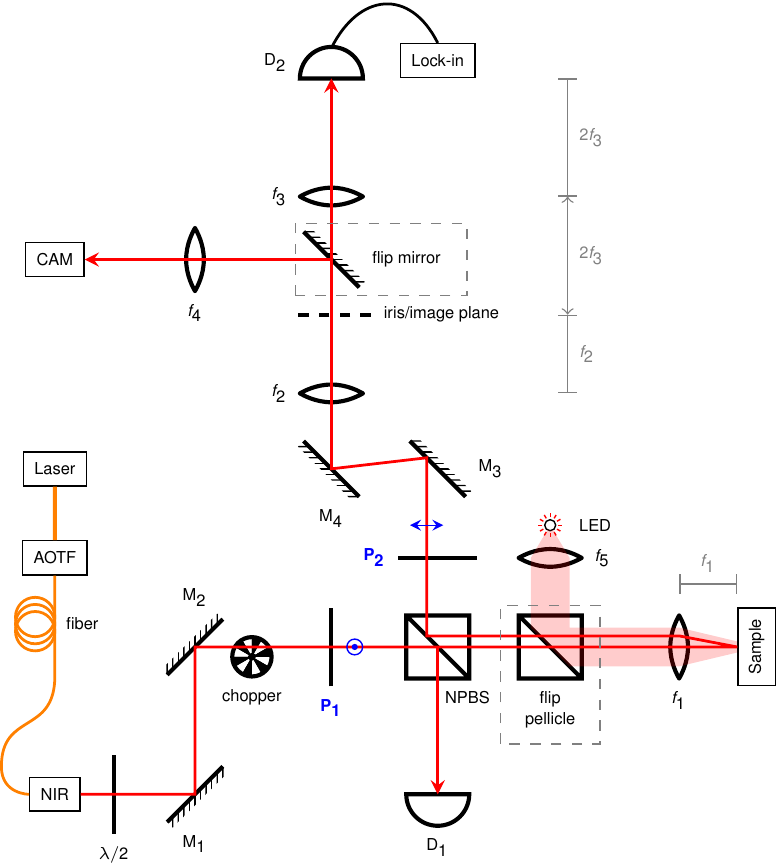}
	\caption{Sketch of the optical measurement setup. Components enclosed in dashed rectangles are removed during measurements.}
	\label{figsetup}
\end{figure}

\section{Optical Measurement Setup}
We use the cross-polarization based measurement method introduced in \cite{Wen2011} to characterize the samples. The dipole antennas emit and receive light in a linearly polarized manner. We shine laser light polarized parallel to the input antenna and collect the light from the output antenna which is rotated 90\textdegree\ with respect to the input antenna [Fig.\ \ref{figsem}(a)]. We suppress the leakage from the input to the output through the use of a crossed polarizer in the output beam path, hence increase the signal-to-noise ratio. We borrow the dipole antenna design from \cite{Kriesch2013} and use a radius of curvature of 3 \textmu{m} for the 90\textdegree\ bends with negligible radiation loss \cite{Kriesch2013}.

The sketch of the optical setup used in the measurements is provided in Fig.\ \ref{figsetup}. Our setup is a modified version of the ones used in \cite{Wen2011,Kriesch2013,Lee2014a}, primarily due to the lack of an infrared InGaAs \textsc{ccd} camera. We use a vidicon infrared camera Electrophysics Micronviewer 7290A (\textsc{cam}) to locate the samples and align the laser beam to the input antenna. We then use a single pixel, 2 mm diameter Thorlabs DET20C InGaAs detector ($D_2$) to measure the light from the output antenna. We also use a Thorlabs S122C Ge detector ($D_1$) to account for any fluctuations in the laser power level.

We use a supercontinuum source (Fianium SC 450-4) connected to an acousto-optic tunable filter (\textsc{aotf}) to get laser light in the 1200--1700 nm wavelength range. The \textsc{aotf} has a long-pass filter installed before its fiber output (Thorlabs FELH1000). Two mirrors ($M_{1,2}$) are used to align the laser beam to the high \textsc{na} microscope objective lens (Leica HC PL FLUOTAR 100x/0.90 POL) with $f_1=2$ mm, used to focus light onto the input antenna. The polarizer $P_1$ (Thorlabs LPNIR050-MP2) is set parallel to the input antenna whereas $P_2$ is set parallel to the output antenna. A Thorlabs BS015 non-polarizing beam splitter (\textsc{npbs}) is used to ensure that the polarization state of the light from the output antenna is preserved. An achromatic doublet (Thorlabs AC254-200-C) with $f_2=200$ mm is used to form an intermediary image plane where magnification is $f_2/f_1=100$. Mirrors $M_{3,4}$ position the light from the output antenna at the center of the intermediary image plane, and an iris is used to filter out all the light except that from the output antenna. We used a removable reticle (Thorlabs R1DS2P) at the image plane to help with the alignment. Another achromatic doublet (Thorlabs AC254-100-C) with $f_3=100$ mm is used to relay the image to $D_2$. Similarly, the lens with $f_4=75$ mm (Thorlabs AC254-075-C) is used to get an overall 150--200x magnification at the camera. A 3W red \textsc{led} (Edison ES S1) is used to illuminate the sample surface with the help of a pellicle beam splitter. 

Our measurement protocol is as follows. We mount the sample on a piezo stage (Thorlabs NanoMax-TS) with pitch and roll control (Thorlabs APR001) and make sure that the sample surface is perpendicular to the laser beam. We set the wavelength to 1550 nm, rotate the Thorlabs AHWP05M-­1600 half-wave plate ($\lambda/2$) and maximize the power through $P_1$. We set $P_2 \parallel P_1$, and focus the laser light on the quartz substrate by minimizing the beam width via adjusting the sample distance to the objective lens by the piezo controller. We then set $P_2 \perp P_1$ and observe the formation of the clover pattern on quartz, typical of Gaussian beams when focused by a high \textsc{na} objective \cite{Kriesch2013}. We move the clover pattern over the input antenna, observe some light from the output antenna, use the iris to block the clover pattern and get light only from the output antenna. We turn-off the \textsc{led}, remove the pellicle, switch to $D_2$, and get a reading from the lock-in amplifier (SRS SR830). We maximize the lock-in reading by moving the sample via the piezo controller in the plane perpendicular to the laser light, without changing the sample to objective lens distance. We then scan the wavelength and record $D_{1,2}$ readings.

\section{Results, Discussion and Conclusion}
\begin{figure}[!t]
\centering
\includegraphics{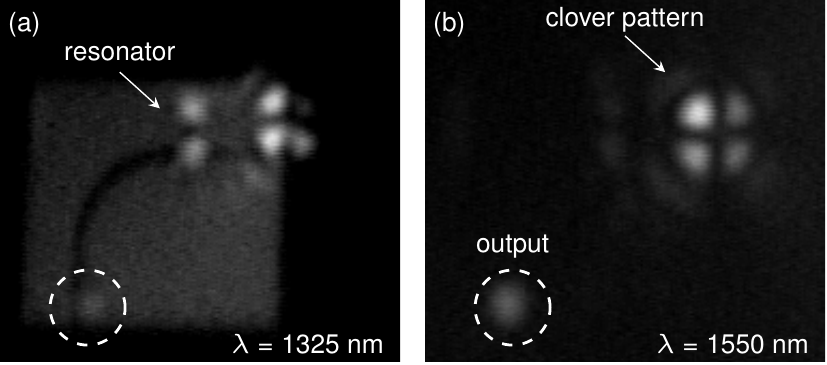}
\caption{A double-stub resonator sample with stub lengths $L_1=L_2=0.98$ \textmu{m} (a) excited at $\lambda=1325$ nm  and imaged with the \textsc{led} light on. (b) The same sample excited at $\lambda=1550$ nm and imaged without \textsc{led} lighting and with the pellicle removed.}
\label{figcam}
\end{figure}

\begin{figure}[!t]
	\centering
	\includegraphics{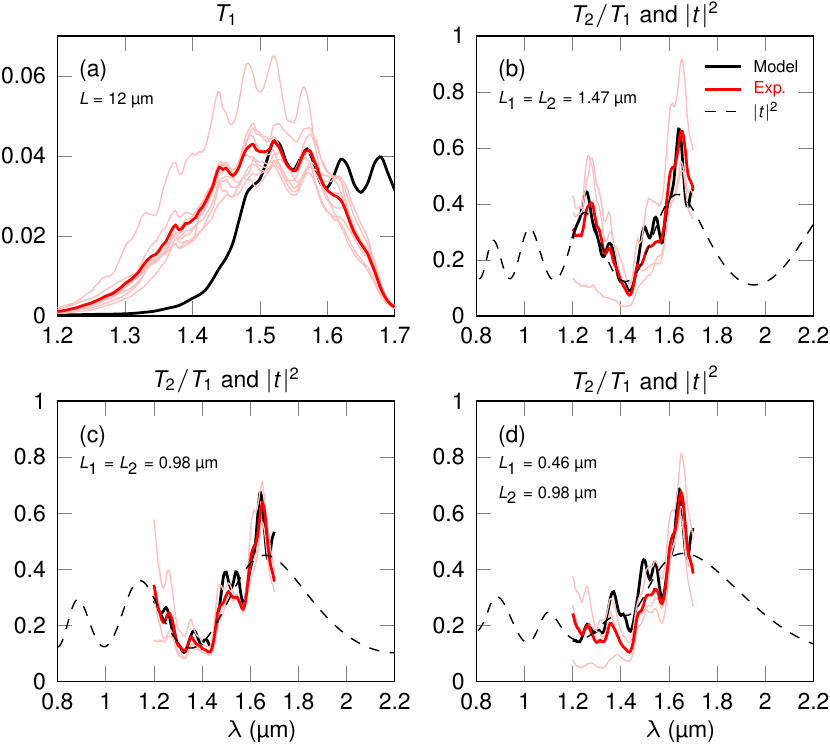}
	\caption{(a) Transmission measurements from nine samples without a resonator (thin pink), their average (thick red) and the estimate from \eqref{eq.T1} (thick black).   (b--d) Transmission from samples with resonators for three different sets of $L_{1,2}$ normalized by $T_1$, from experimental measurements (thin pink), their average (thick red), estimate from \eqref{eq.T1}--\eqref{eq.T2} (thick black) and plot of $\lvert t\rvert^2$ (dashed). Fitted $L_{1,2}$ values are shown in (b--d).}
	\label{figdata}
\end{figure}

We characterized a number of different samples with and without resonators. Fig.\ \ref{figcam}(a) shows the captured camera image\footnote{The fact that the illumination \textsc{led} is at $\sim$630 nm and the laser light is at 1325 nm leads to a small shift and defocus of the sample image on the camera with respect to the laser beams due to the chromatic aberrations in the system.} for a double-stub resonator sample when the stub is resonant and has a large $\lvert r\rvert$ [equivalently $\lvert t\rvert$ is near its minimum, see dashed line in Fig.\ \ref{figdata}(c)]. There is considerable leakage from the stub section, and the antenna output intensity is quite low, as highlighted by the white-dashed circle. On the other hand, when the wavelength moves away from the stub resonance and $\lvert t\rvert$ increases, leakage from the stub decreases and the antenna output becomes more pronounced [Fig.\ \ref{figcam}(b)]. 

We show the results for the power transmission through waveguides without stub resonators ($T_1$) in Fig.\ \ref{figdata}(a). We normalize the experimental curves with an estimate of the incident power which is obtained by measuring the reflection of laser light from a large metal patch on the sample while $P_2 \parallel P_1$. We obtain data from nine different samples, similar to Fig.\ \ref{figsem}(a), with a total waveguide length $L=12$ {\textmu}m. Although the nine samples had exactly the same e-beam mask, due to variations during the fabrication, we see variations in $T_1$. We average over the measurements to obtain the thick red line in Fig.\ \ref{figdata}(a). We also calculate $T_1$ via \eqref{eq.T1}, where in $S_{12}$ simulations we take into account the changes in the Gaussian beam width and center position as the wavelength is scanned (measured experimentally from camera images of the beam reflecting off of a blank quartz section on the sample). We normalize the calculated $T_1$ (black curve) so that it has the same maximum value as the average measurement results (red curve). The fringes in $T_1$ are due to the low-Q cavity formed by the antennas and the $L=12$ {\textmu}m waveguide. As reported in \cite{Ly-Gagnon2012} the fringe spacing and depth can be used to obtain the real and imaginary parts of $k$, respectively. The calculated $k$ values correctly gave us the fringe spacing which is a function of the real part of $k$; to match the fringe depths, we had to increase the imaginary part of $k$ by 50\% to give us a power transmission length, $\frac{1}{2\Imag(k)}$, of 5.3 {\textmu}m for the waveguide mode at 1550 nm. We get very good correspondence between the measurements and the model around 1550 nm, the wavelength at which we focus our laser beam. The correspondence deteriorates at lower and higher wavelengths, probably due to the non-orthogonality of the sample surface to the laser beam and chromatic aberrations.

We measure the power transmitted through samples with stub resonators ($T_2$) for three different sets of stub lengths ($L_{1,2}$) from samples similar to Fig.\ \ref{figsem}(b). We plot measurement results for $T_2/T_1$ for each $L_{1,2}$ set in Fig.\ \ref{figdata}(b--d) where $T_1$ is taken to be the average measurement in Fig.\ \ref{figdata}(a). In each subplot, we have three different samples with the same e-beam mask dimensions (thin pink lines) with variations in the measurements due to fabrication. We plot the average of the experimental measurements with thick red lines. We also plot calculated $T_2/T_1$ from \eqref{eq.T1}--\eqref{eq.T2} (thick black lines) and calculated $\lvert t \rvert^2$ from \cite{Naghizadeh2017} (black dashed lines). The e-beam mask dimensions for the stubs were $L_{1,2}=1380$ nm for Fig.\ \ref{figdata}(b), $L_{1,2}=910$ nm for Fig.\ \ref{figdata}(c), and $L_1=430, L_2 = 910$ nm for Fig.\ \ref{figdata}(d). These values were chosen to have a transmission maximum at 1550 nm \cite{Naghizadeh2017}. However, we had longer stubs in the fabricated samples as observed from \textsc{sem} images, and the resonators were all resonant at 1650 nm. We fitted $L_{1,2}$ values so that the experimental $T_2/T_1$ values and those calculated via \eqref{eq.T1}--\eqref{eq.T2} agree well with each other. In \eqref{eq.T1}--\eqref{eq.T2} we use the $k$ values obtained from Fig.\ \ref{figdata}(a). The fitted values of $L_{1,2}$ reported in Fig.\ \ref{figdata}(b--d) are ${\sim}6\%$ larger than those on the e-beam mask. The fitted values are also used in the calculation of $\lvert t \rvert^2$. As can be seen, there is a good overlap between the predictions of the scattering matrix model and the experimental data. Furthermore, we see that $T_2/T_1$ follows $\lvert t \rvert^2$, with extra ripples due to the low-Q resonances set up in between the two antennas and the connecting waveguide.

In conclusion, we experimentally characterized stub resonators embedded in plasmonic slot waveguides. We used the scattering matrix based characterization of the resonators in an extended model which takes into account the coupling of a Gaussian beam into the slot waveguide mode by an optical antenna. We provided details of the optical characterization setup and have shown that the measurements and modeling agree with each other. Our results are also relevant for stub designs at THz frequencies \cite{Tao2013}. The development of compact resonators embedded in plasmonic slot waveguides can pave the way to sensors with an improved figure of merit as well as detectors with very low energy consumption. 


\begin{thebibliography}{10}
\providecommand{\url}[1]{#1}
\csname url@samestyle\endcsname
\providecommand{\newblock}{\relax}
\providecommand{\bibinfo}[2]{#2}
\providecommand{\BIBentrySTDinterwordspacing}{\spaceskip=0pt\relax}
\providecommand{\BIBentryALTinterwordstretchfactor}{4}
\providecommand{\BIBentryALTinterwordspacing}{\spaceskip=\fontdimen2\font plus
\BIBentryALTinterwordstretchfactor\fontdimen3\font minus
  \fontdimen4\font\relax}
\providecommand{\BIBforeignlanguage}[2]{{%
\expandafter\ifx\csname l@#1\endcsname\relax
\typeout{** WARNING: IEEEtran.bst: No hyphenation pattern has been}%
\typeout{** loaded for the language `#1'. Using the pattern for}%
\typeout{** the default language instead.}%
\else
\language=\csname l@#1\endcsname
\fi
#2}}
\providecommand{\BIBdecl}{\relax}
\BIBdecl

\bibitem{Miller2017}
D.~A.~B. Miller, ``\href{http://dx.doi.org/10.1109/JLT.2017.2647779}{Attojoule
  optoelectronics for low-energy information processing and communications -- a
  tutorial review},'' \emph{Journal of Lightwave Technology}, vol.~PP, no.~99,
  pp. 1--1, 2017, (early access).

\bibitem{Kinsey2015}
N.~Kinsey, M.~Ferrera, V.~M. Shalaev, and A.~Boltasseva,
  ``\href{http://dx.doi.org/10.1364/JOSAB.32.000121}{Examining nanophotonics
  for integrated hybrid systems: a review of plasmonic interconnects and
  modulators using traditional and alternative materials},'' \emph{J. Opt. Soc.
  Am. B}, vol.~32, no.~1, pp. 121--142, Jan 2015.

\bibitem{Veronis2007a}
G.~Veronis and S.~H. Fan,
  ``\href{http://dx.doi.org/10.1109/JLT.2007.903544}{Modes of subwavelength
  plasmonic slot waveguides},'' \emph{Journal of Lightwave Technology},
  vol.~25, no.~9, pp. 2511--2521, September 2007.

\bibitem{Wen2011}
J.~Wen, P.~Banzer, A.~Kriesch, D.~Ploss, B.~Schmauss, and U.~Peschel,
  ``\href{http://dx.doi.org/DOI:10.1063/1.3564904}{Experimental
  cross-polarization detection of coupling far-field light to highly confined
  plasmonic gap modes via nanoantennas},'' \emph{Applied Physics Letters},
  vol.~98, no.~10, p. 101109, 2011.

\bibitem{Andryieuski2014}
A.~Andryieuski, V.~A. Zenin, R.~Malureanu, V.~S. Volkov, S.~I. Bozhevolnyi, and
  A.~V. Lavrinenko, ``\href{http://dx.doi.org/10.1021/nl501207u}{Direct
  characterization of plasmonic slot waveguides and nanocouplers},'' \emph{Nano
  Lett.}, vol.~14, no.~7, pp. 3925--3929, Jul. 2014.

\bibitem{Gao2016a}
Q.~Gao, F.~Ren, and A.~X. Wang,
  ``\href{http://dx.doi.org/10.1109/LPT.2016.2533583}{Direct and efficient
  optical coupling into plasmonic integrated circuits from optical fibers},''
  \emph{IEEE Photonics Technology Letters}, vol.~28, no.~11, pp. 1165--1168,
  June 2016.

\bibitem{Fu2012}
Y.~Fu, X.~Hu, C.~Lu, S.~Yue, H.~Yang, and Q.~Gong,
  ``\href{http://dx.doi.org/10.1021/nl303095s}{All-optical logic gates based on
  nanoscale plasmonic slot waveguides},'' \emph{Nano Letters}, vol.~12, no.~11,
  pp. 5784--5790, 2012.

\bibitem{Ly-Gagnon2012}
D.-S. Ly-Gagnon, K.~C. Balram, J.~S. White, P.~Wahl, M.~L. Brongersma, and
  D.~A.~B. Miller, ``\href{http://dx.doi.org/10.1515/nanoph-2012-0002}{Routing
  and photodetection in subwavelength plasmonic slot waveguides},''
  \emph{Nanophotonics}, vol.~1, no.~1, pp. 9--16, 2012.

\bibitem{Geisler2013}
P.~Geisler, G.~Razinskas, E.~Krauss, X.-F. Wu, C.~Rewitz, P.~Tuchscherer,
  S.~Goetz, C.-B. Huang, T.~Brixner, and B.~Hecht,
  ``\href{http://dx.doi.org/10.1103/PhysRevLett.111.183901}{Multimode plasmon
  excitation and \textit{In~Situ} analysis in top-down fabricated
  nanocircuits},'' \emph{Phys. Rev. Lett.}, vol. 111, p. 183901, Oct 2013.

\bibitem{Dai2014}
W.-H. Dai, F.-C. Lin, C.-B. Huang, and J.-S. Huang,
  ``\href{http://dx.doi.org/10.1021/nl501102n}{Mode conversion in
  high-definition plasmonic optical nanocircuits},'' \emph{Nano Lett.},
  vol.~14, no.~7, pp. 3881--3886, Jul. 2014.

\bibitem{Kriesch2013}
A.~Kriesch, S.~P. Burgos, D.~Ploss, H.~Pfeifer, H.~A. Atwater, and U.~Peschel,
  ``\href{http://dx.doi.org/10.1021/nl402580c}{Functional plasmonic
  nanocircuits with low insertion and propagation losses},'' \emph{Nano
  Letters}, vol.~13, no.~9, pp. 4539--4545, 2013.

\bibitem{Rewitz2014}
C.~Rewitz, G.~Razinskas, P.~Geisler, E.~Krauss, S.~Goetz, M.~Paw\l{}owska,
  B.~Hecht, and T.~Brixner,
  ``\href{http://dx.doi.org/10.1103/PhysRevApplied.1.014007}{Coherent control
  of plasmon propagation in a nanocircuit},'' \emph{Phys. Rev. Applied},
  vol.~1, p. 014007, Feb 2014.

\bibitem{Melikyan2014}
A.~Melikyan, L.~Alloatti, A.~Muslija, D.~Hillerkuss, P.~C. Schindler, J.~Li,
  R.~Palmer, D.~Korn, S.~Muehlbrandt, D.~V. Thourhout, B.~Chen, R.~Dinu,
  M.~Sommer, C.~Koos, M.~Kohl, W.~Freude, and J.~Leuthold,
  ``\href{http://dx.doi.org/10.1038/NPHOTON.2014.9}{High-speed plasmonic phase
  modulators},'' \emph{Nat Photon}, vol.~8, no.~3, pp. 229--233, Mar. 2014.

\bibitem{Lee2014a}
H.~W. Lee, G.~Papadakis, S.~P. Burgos, K.~Chander, A.~Kriesch, R.~Pala,
  U.~Peschel, and H.~A. Atwater,
  ``\href{http://dx.doi.org/10.1021/nl502998z}{Nanoscale conducting oxide
  {PlasMOStor}},'' \emph{Nano Letters}, vol.~14, no.~11, pp. 6463--6468, 2014.

\bibitem{No2013}
Y.-S. No, J.-H. Choi, H.-S. Ee, M.-S. Hwang, K.-Y. Jeong, E.-K. Lee, M.-K. Seo,
  S.-H. Kwon, and H.-G. Park, ``\href{http://dx.doi.org/10.1021/nl3044822}{A
  double-strip plasmonic waveguide coupled to an electrically driven nanowire
  \textsc{led}},'' \emph{Nano Lett.}, vol.~13, no.~2, pp. 772--776, Feb. 2013.

\bibitem{Huang2014}
K.~C.~Y. Huang, M.-K. Seo, T.~Sarmiento, Y.~Huo, J.~S. Harris, and M.~L.
  Brongersma, ``\href{http://dx.doi.org/10.1038/NPHOTON.2014.2}{Electrically
  driven subwavelength optical nanocircuits},'' \emph{Nat Photon}, vol.~8,
  no.~3, pp. 244--249, Mar. 2014.

\bibitem{Bozhevolnyi2006}
S.~I. Bozhevolnyi, V.~S. Volkov, E.~Devaux, J.-Y. Laluet, and T.~W. Ebbesen,
  ``\href{http://dx.doi.org/10.1038/nature04594}{Channel plasmon subwavelength
  waveguide components including interferometers and ring resonators},''
  \emph{Nature}, vol. 440, no. 7083, pp. 508--511, Mar. 2006.

\bibitem{Zhu2014}
Y.~Zhu, X.~Hu, H.~Yang, and Q.~Gong,
  ``\href{http://dx.doi.org/10.1038/srep03752}{On-chip plasmon-induced
  transparency based on plasmonic coupled nanocavities},'' \emph{Scientific
  Reports}, vol.~4, p. 3752, Jan. 2014.

\bibitem{Chai2016}
Z.~Chai, X.~Hu, C.~Li, H.~Yang, and Q.~Gong,
  ``\href{http://dx.doi.org/10.1021/acsphotonics.6b00399}{On-chip multiple
  electromagnetically induced transparencies in photon--plasmon composite
  nanocavities},'' \emph{ACS Photonics}, Oct. 2016.

\bibitem{Raza2016}
S.~Raza, M.~Esfandyarpour, A.~L. Koh, N.~A. Mortensen, M.~L. Brongersma, and
  S.~I. Bozhevolnyi, ``\href{http://dx.doi.org/10.1038/ncomms13790}{Electron
  energy-loss spectroscopy of branched gap plasmon resonators},'' \emph{Nature
  Communications}, vol.~7, p. 13790, Dec. 2016.

\bibitem{Naghizadeh2017}
S.~Naghizadeh and \c{S}\"{u}kr\"{u} Ekin~Kocaba\c{s},
  ``\href{http://dx.doi.org/10.1364/JOSAB.34.000207}{Guidelines for designing
  {2D} and {3D} plasmonic stub resonators},'' \emph{J. Opt. Soc. Am. B},
  vol.~34, no.~1, pp. 207--217, Jan 2017.

\bibitem{McPeak2015}
K.~M. McPeak, S.~V. Jayanti, S.~J.~P. Kress, S.~Meyer, S.~Iotti, A.~Rossinelli,
  and D.~J. Norris, ``\href{http://dx.doi.org/10.1021/ph5004237}{Plasmonic
  films can easily be better: Rules and recipes},'' \emph{ACS Photonics},
  vol.~2, no.~3, pp. 326--333, 2015.

\bibitem{Kocabas2008}
\c{S}. E.~Kocaba\c{s}, G.~Veronis, D.~A.~B. Miller, and S.~Fan,
  ``\href{http://dx.doi.org/10.1109/JSTQE.2008.924431}{Transmission line and
  equivalent circuit models for plasmonic waveguide components},''
  \emph{Selected Topics in Quantum Electronics, IEEE Journal of}, vol.~14,
  no.~6, pp. 1462--1472, Nov.-Dec. 2008.

\bibitem{Tao2013}
J.~Tao, B.~Hu, X.~Y. He, and Q.~J. Wang,
  ``\href{http://dx.doi.org/10.1109/TNANO.2013.2285127}{Tunable subwavelength
  terahertz plasmonic stub waveguide filters},'' \emph{IEEE Transactions on
  Nanotechnology}, vol.~12, no.~6, pp. 1191--1197, Nov 2013.

\end{thebibliography}

\end{document}